\documentstyle[preprint,aps,epsf]{revtex}
\tightenlines
\begin{document}
\draft
\title{
%
%
\begin{flushright}
\normalsize
KEK-CP-072          \\
KEK Preprint 97-218 \\
November 1997       \\
\end{flushright}
%
%
$K^{+}\to\pi^{+}\pi^{0}$ Decay Amplitude with the Wilson Quark Action
in Quenched Lattice QCD
}
%
\author{
JLQCD Collaboration \\
      S.~Aoki$^{\rm a }$\footnote{
                           present address: Max-Planck-Institut f\"ur Physik,
                           F\"ohringer Ring 6, D-80805 M\"unchen, Germany.},
  M.~Fukugita$^{\rm b }$,
 S.~Hashimoto$^{\rm c }$\footnote{
                           present address: Theoretical Physics Department,
                           Fermi National Accelerator Laboratory, P.O. Box 500,
                           Batavia, IL 60510, USA.},
  N.~Ishizuka$^{\rm a, d }$,
   Y.~Iwasaki$^{\rm a, d }$,
    K.~Kanaya$^{\rm a, d }$,
 Y.~Kuramashi$^{\rm e }$,
     M.~Okawa$^{\rm e }$,
     A.~Ukawa$^{\rm a }$,
T.~Yoshi\'{e}$^{\rm a, d }$
}
\address{
${}^{\rm a}$Institute of Physics, University of Tsukuba, Tsukuba, Ibaraki 305, Japan \\
${}^{\rm b}$Institute for Cosmic Ray Research, University of Tokyo, Tanashi, Tokyo 188, Japan \\
${}^{\rm c}$Computing Research Center, High Energy Accelerator Research Organization~(KEK), Tsukuba, Ibaraki 305, Japan \\
${}^{\rm d}$Center for Computational Physics, University of Tsukuba, Tsukuba, Ibaraki 305, Japan \\
${}^{\rm e}$Institute of Particle and Nuclear Studies, High Energy Accelerator Research Organization~(KEK), Tsukuba, Ibaraki 305, Japan \\
}
\date{\today}
\maketitle
%
%
\begin{abstract}
We present a calculation for the $K^+\to\pi^+ \pi^0 $ decay amplitude
using a quenched simulation of lattice QCD with the Wilson quark
action at $\beta=6/g^2=6.1$.
The decay amplitude is extracted from the ratio, the
$K\to\pi\pi$ three-point function divided by either
$K$ and $\pi$ meson two-point functions or
$K$ meson two-point function and $I=2$ $\pi\pi$ four-point function;
the two different methods yield consistent results.
Finite size effects are examined with calculations made on $24^3\times 64$
and $32^3\times 64$ lattices, and are shown that they are explained by one-loop
effects of chiral perturbation theory.
The lattice amplitude is converted to the continuum value
by employing a one-loop calculation of chiral
perturbation theory, yielding
a value in agreement with
experiment if extrapolated to the chiral limit.
We also report on the $K$ meson $B$ parameter $B_K$ obtained from
the $K^+\to\pi^+\pi^0$ amplitude using chiral perturbation theory.
\end{abstract}
%
%
\narrowtext
\section{ Introduction }
Despite the full understanding of the fundamental theory of weak interactions,
the non-leptonic decay of hadrons still remains to be the least understood
weak process, the most notable problems being $\Delta I=1/2$ rule and
the calculation of $\epsilon^\prime/\epsilon$.
The predicament originates from the difficulty of evaluating
the hadronic matrix element of the product of currents.
Much work has already been done to attack this problem
using lattice QCD simulations\cite{ELConehalf,BSKpipi,LANLonehalf,KPPE-GA},
but they have not yielded satisfactory results.

Difficulties have proven to be especially severe for the
the $\Delta I=1/2$ amplitudes\cite{ELConehalf,BSKpipi,LANLonehalf}.
From the computational point of view the problem lies in a calculation of the
so-called eye diagram, which suffers from extremely large statistical
fluctuations\cite{ELConehalf,BSKpipi,LANLonehalf,JLQCD}.
Theoretically, this may be related to mixing of the
dimension six weak operator responsible for the decay with
operators of lower dimensions whose coefficients diverge linearly in the continuum limit.
At a more fundamental level, there is the difficulty \cite{mainitesta}
that the $K\to\pi\pi$ 3-point function evaluated in Euclidean space-time
does not yield information on the phase of the decay amplitude.

Calculation of the $\Delta I=3/2$ process is known to be
easier than that of the $\Delta I=1/2$ process.
For this case mixing of lower dimension operators is absent,
and the so-called figure-eight diagrams which represent
the $\Delta I=3/2$ amplitude, have clear signals in numerical simulations.
Indeed lattice calculations have been reported by several
groups\cite{BSKpipi,KPPE-GA} quite a long time ago.
The problem, however, was that the results turned out to be
inconsistent with experiment: lattice calculations have given
the amplitude roughly a factor of two larger than experiment.

Two potential origins are suspected to give this discrepancy.
One is an issue in the matching of the lattice and continuum operators.
Early studies employed the factor $\sqrt{2\kappa}$
for the quark wave function normalization and the bare lattice
coupling constant for estimating
the renormalization factor of the four-quark weak operator.
It is by now well known that the KLM factor
$\sqrt{1-3\kappa/4\kappa_c}$\cite{KLM}
and tadpole-improved perturbation theory\cite{TPI} are more adequate
for the operator matching.

Another problem concerns the use of chiral perturbation theory (CHPT)
to convert lattice results into the physical amplitude.
Only the tree-level formula was known and used in the previous work.
The meson mass dependence of lattice calculations
appeared consistent with the prediction of the tree-level formula,
allowing, however, for large statistical errors.
It was probably necessary to use the formula including higher order
CHPT effects, but its necessity was not manifest.
An interesting development in this connection is a recent calculation
of one-loop corrections to the $K^+\to\pi^+\pi^0$ amplitude in CHPT by
Golterman and Leung\cite{ONECPTH}.
Applying their results to the old data obtained by Bernard and Soni\cite{BSKpipi},
they found that one-loop effects decrease the physical amplitude by about 30\%.

With the hope to improve the problems posed here,
we have carried out a high statistics simulation of the $K^+\to\pi^+\pi^0$
amplitude in quenched lattice QCD,
incorporating various theoretical and technical developments made
in recent years.
In particular, we discuss in detail how one-loop corrections of CHPT affect
physical predictions for the decay amplitude from lattice QCD simulations.
We also report on the $K$ meson $B$ parameter $B_K$ obtained from the
$K^+\to\pi^+\pi^0$ amplitude using CHPT.

This paper is organized as follows.
After a brief description of simulation parameters in Sec.~\ref{TWO_S},
we explain our method for extracting the decay amplitude in Sec.~\ref{THREE_S}.
Our results for the $K^+\to\pi^+\pi^0$ amplitude are presented in Sec.~\ref{FOUR_S}
with discussion made on one-loop effects of CHPT.
Results for $B_K$ are given in Sec.~\ref{FIVE_S}.
Sec.~\ref{SIX_S} summarizes our conclusions.
%
%
\section{\label{TWO_S} Simulation parameters }
Our simulation is carried out in quenched lattice QCD employing
the standard plaquette action for gluons at $\beta=6.1$ and the Wilson action for quarks.
We take up, down and strange quarks to be degenerate, and make measurements
at four values of the common hopping parameter,
$\kappa = 0.1520$, $0.1530$, $0.1540$, and $0.1543$, which correspond to
$M_\pi/M_\rho = 0.797$, $0.734$, $0.586$ and $0.515$.
In order to examine finite-size effects, simulations are carried out for
two lattice sizes, 120 configurations on $24^3\times 64$ and 65 configurations
on $32^3\times 64$. Gluon configurations are separated by $2000$ pseudo heat bath sweeps.
Quark propagators are solved with the Dirichlet boundary condition imposed in the time direction
and the periodic boundary condition in the space directions.

We adopt $1/a=2.67(10){\rm GeV}$ for the physical scale of lattice spacing
estimated from the $\rho$ meson mass, and $\kappa_c=0.15499(2)$
for the critical hopping parameter, which were obtained in our
previous study\cite{JLQCDfB}.
Our calculations are carried out on the Fujitsu VPP500/80 supercomputer at KEK.
%
%
\section{\label{THREE_S} Methods }
\subsection{ Extraction of decay amplitude }
Let us consider the four-quark operator defined by
\begin{equation}
Q_+ = \frac{1}{2} \cdot
      [ \ \bar{s} \gamma_\mu ( 1 - \gamma_5 ) d
        \ \bar{u} \gamma_\mu ( 1 - \gamma_5 ) u
  +       \bar{s} \gamma_\mu ( 1 - \gamma_5 ) u
	\ \bar{u} \gamma_\mu ( 1 - \gamma_5 ) d \ ] \ ,
\label{QP}
\end{equation}
which is relevant to $\Delta I=3/2$ two-pion decay of the $K$ meson.
We first discuss our method for extracting the lattice matrix element of the operator $Q_+$,
deferring the question of matching the lattice and continuum operators
to Sec.~\ref{OperatorMatching}.

We extract the decay amplitude from the 4-point correlation function
\begin{equation}
  G_Q ( t_+, t_0 ; t ;  t_K )
= \langle 0 | W_+ (t_+) W_0(t_0) Q_{+}(t) W_K (t_K) | 0 \rangle \ .
\end{equation}
In order to enhance signals we construct wall sources (denoted by $W$)
for all external mesons, and fix gauge configurations to the Coulomb gauge.
The wall sources $W_K$, $W_+$, and $W_0$ for $K^{+}$, $\pi^{+}$, and $\pi^{0}$
are placed at the time slices $t_K$, $t_+$ and $t_0$ such that
$t_K\ll t\ll t_+, t_0$.
All mesons are at rest, and the 4-quark operator $Q_+$ is
projected to zero spatial momentum.

In our calculation for temporal lattice size $T=64$, we place the $K$ meson at $t_K=4$.
The two $\pi$ mesons are placed at different time slices, $t_+=59$ and $t_0=60$
to avoid contaminations from Fierz-rearranged terms in the two-pion state
that would occur for the choice $t_+=t_0$.

The correlation function $G_Q$ behaves for $t_K \ll t \ll t_+ \sim t_0$ as
\begin{eqnarray}
&&   G_Q ( t_+, t_0 ; t ; t_K ) \nonumber\\
&& =
     \langle 0           | W_+(0) W_0(t_0-t_+) | \pi^+ \pi^0 \rangle { 1 \over N_{\pi\pi} }
     \langle \pi^+ \pi^0 | Q_+(0)              | K^+         \rangle { 1 \over N_K        }
     \langle K^+         | W_K(0)              | 0           \rangle \nonumber \\
&&   \quad
     \times e^{ M_K (t_K-t) } \cdot e^{(t - t_+) M_{\pi\pi}}
\label{MQ}
\end{eqnarray}
where $N_K$ denotes the normalization factor of the $K$ meson state,
$| \pi^+ \pi^0 \rangle$ represents the $I=2$ two-pion state with a mass
$M_{\pi\pi}$ and a state normalization factor $N_{\pi\pi}$.

In order to remove the normalization factors in $G_Q$
we calculate the product of the meson 2-point functions given by
\begin{equation}
  G_W ( t_+, t_0 ; t ; t_K ) =
                   \langle 0 | W_0 (t_0) \pi^{0} ( t ) | 0 \rangle
                   \langle 0 | W_+ (t_+) \pi^{+} ( t ) | 0 \rangle
                   \langle 0 | K^+ (t)   W_K (t_K)     | 0 \rangle  \ .
\end{equation}
Defining a ratio $R_W = G_Q / G_W$, we find
\begin{equation}
  R_W ( t_+, t_0 ; t ; t_K ) = S_W \cdot
    \frac{ \langle \pi^+ \pi^0 | Q_+ | K^+ \rangle }{ \langle \pi | \pi | 0 \rangle^3 }
    \cdot {\rm e}^{ (t-t_+) \Delta }
\end{equation}
where
\begin{equation}
    \Delta = M_{\pi\pi}- 2 M_\pi
\end{equation}
is a mass shift due to a finite spatial lattice size,
and $S_W$ is defined by
\begin{equation}
   S_W =
        \frac{ N_\pi^2 }{ N_{\pi\pi} }
     \cdot
        \frac{ \langle 0 | W_+(0) W_0(t_0-t_+) | \pi^+ \pi^0 \rangle }
             { \langle 0 | W_+(0)              | \pi^+       \rangle
               \langle 0 | W_0(t_0-t_+)        | \pi^0       \rangle }
\end{equation}
where $t_0-t_+=1$ in our calculation.  The value of $S_W$ 
should converge to unity for infinite volume. 

\begin{figure}
\centerline{\epsfxsize=14.0cm \epsfbox{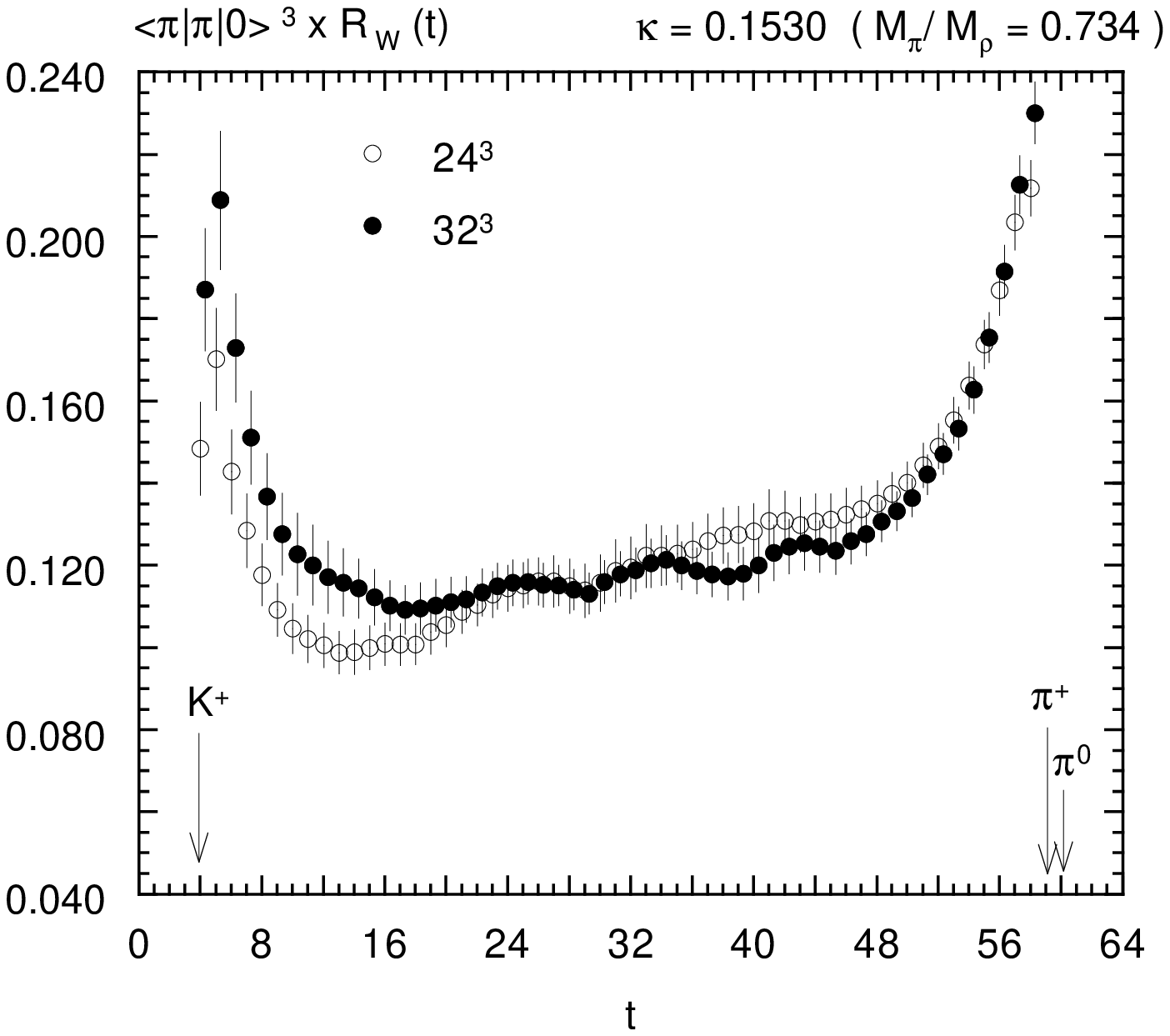}}
\caption{ \label{RWF}
$\langle \pi | \pi | 0 \rangle^3 \cdot R_W( t_+, t_0 ; t ; t_K )$
at $\kappa=0.153$.
Open and filled circles refer to data for $24^3$ and $32^3$ lattices.
}
\end{figure}

In Fig.~\ref{RWF} we plot $\langle \pi | \pi | 0 \rangle^3 \cdot R_W$ at
$\kappa=0.1530$ as a function of time $t$ of the weak operator,
where we calculate $\langle \pi | \pi | 0\rangle$ from the pion 2-point function
for point source and point sink.
We observe a clear non-vanishing slope, which means the mass shift $\Delta$
being positive.
Numerical values of $\Delta$ and the decay amplitude
$\langle \pi^+ \pi^0 | Q_+| K^+ \rangle$ obtained by a single exponential fit
for the time range $t=18-46$ are tabulated in Table~\ref{RWRPT}.
Here we assume $S_W=1$,
whose justification will be discussed below.

According to L\"uscher's formula\cite{LUSHER},
the finite-size mass shift of the two-pion state is written
\begin{equation}
   \Delta = M_{\pi\pi} - 2 M_\pi = - \frac{4\pi a_0}{ M_\pi (a L)^3 } + O(L^{-4})
\end{equation}
where $a_0$ is the $s$-wave scattering length and $L$ is the spatial size.
This formula was previously employed to calculate the s-wave $\pi\pi$
scattering length in quenched lattice QCD\cite{PIPI-SGK,PIPI-GPS,PIPI-K}.
It was found that lattice calculations give $a_0$
in good agreement with the prediction of current algebra.
Using the current algebra formula $a_0^{I=2}=M_\pi / (8\pi F_\pi^2 )$ with
$F_\pi=132{\rm MeV}$ and $1/a=2.67(10){\rm GeV}$,
we obtain $a \Delta=0.015$ for $L=24$ and 0.006 for $L=32$.
Considering uncertainties arising from terms of $O(L^{-4})$ and the
difference between the physical and
measured values of $F_\pi$,
we regard this estimate being consistent with the measured $a\Delta$
(see Table~\ref{RWRPT}).

As an alternative method we may
remove the normalization factors of
the 4-point function $G_Q$  with
\begin{equation}
G_P ( t_+, t_0 ; t ; t_K ) =
    \langle 0 | W_+ (t_+) W_0 (t_0) \pi^{+} ( t ) \pi^{0} ( t ) | 0 \rangle
    \langle 0 | K^+ (t) W_K (t_K)      | 0 \rangle \ .
\end{equation}
The ratio $R_P = G_Q / G_P$ is
independent of $t$ and it does not depend on the wall sources for
$t_K \ll t \ll t_+ \sim t_0$,
\begin{equation}
R_P ( t_+, t_0 ; t ; t_K ) = S_P^{-1} \cdot
        \frac{ \langle \pi^+ \pi^0 | Q_+ | K^+ \rangle }
             { \langle \pi | \pi | 0 \rangle^3 }
\label{RP}
\end{equation}
where
\begin{equation}
S_P =  \frac{ \langle \pi^+ \pi^0  | \pi^+ \pi^0  | 0 \rangle }
            { \langle \pi | \pi | 0 \rangle^2 }
\end{equation}
which should become unity for infinite spatial lattice.

\begin{figure}
\centerline{\epsfxsize=14.0cm \epsfbox{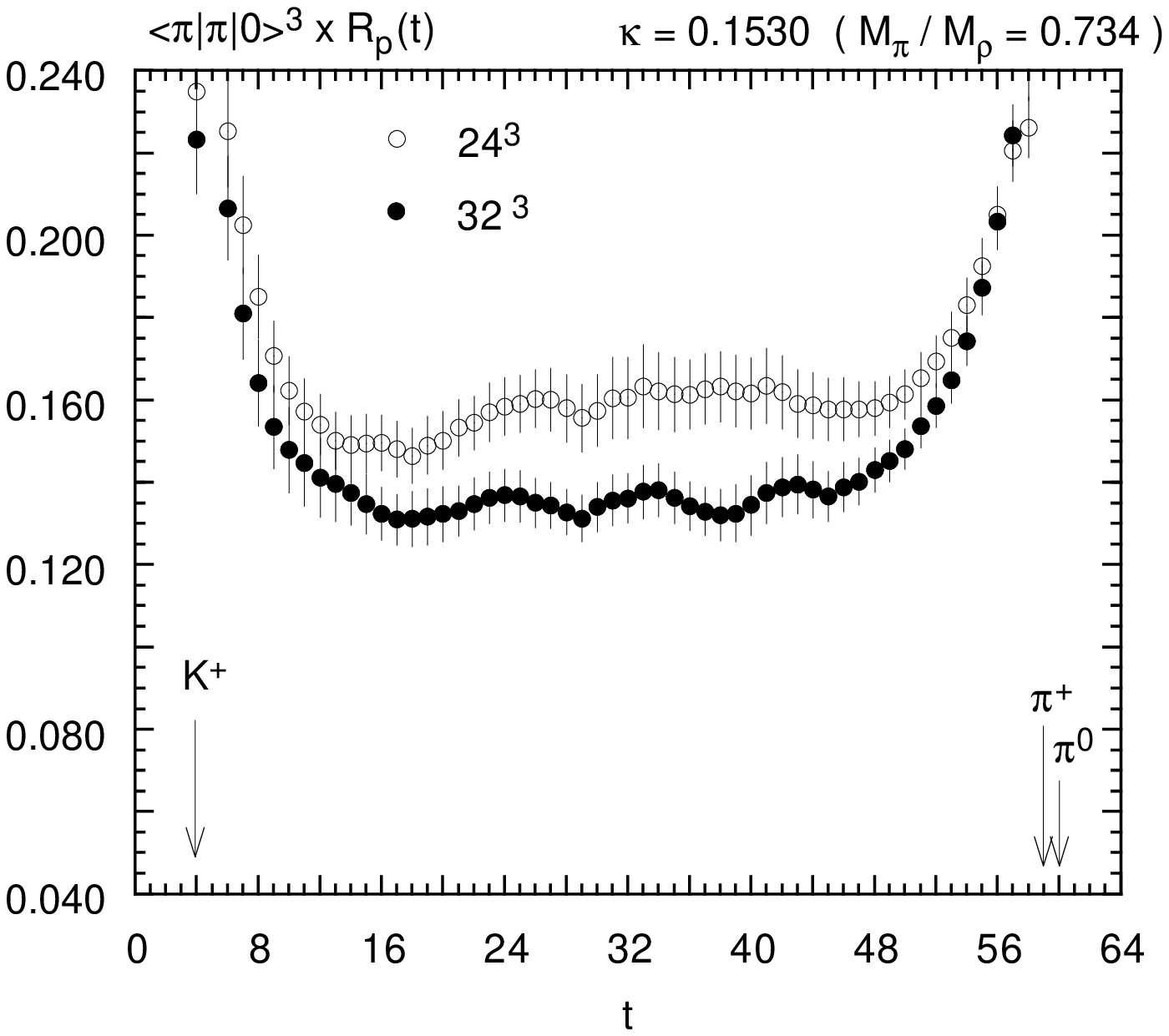}}
\caption{\label{RPF}
$\langle \pi | \pi | 0 \rangle^3 \cdot R_P( t_+, t_0 ; t ; t_K )$
at $\kappa=0.153$.
Open and filled circles refer to data for $24^3$ and $32^3$ lattices.
}
\end{figure}

The dependence of $R_P$ on the time $t$ of the weak operator
is shown in Fig.~\ref{RPF} for $\kappa=0.153$, the same hopping
parameter
as in Fig.~\ref{RWF} for $R_W$.
As expected, a clear plateau is seen for $t\approx 20-40$, where
effects of excited states near the lattice boundaries already disappear.

In Table~\ref{RWRPT} we list
$\langle \pi^+ \pi^0 | Q_+ | K^+ \rangle$ obtained by fitting $R_P$ to a constant
over $t=22-42$ assuming $S_P=1$.
The results from the two methods show good mutual agreement, well
within the statistical error of $10-15$\%.
We note that statistical errors for $R_P$ are smaller, and
therefore adopt the matrix elements from $R_P$ to obtain the physical
decay amplitude below.

We still have to justify the assumption $S_W=S_P=1$ used above.
This is not {\it a priori} obvious, especially for $S_W$,
since wall sources are uniformly extended across the spatial lattice,
although a good agreement of
$\langle \pi^+ \pi^0 | Q_+ | K^+ \rangle$ from $R_W$ and $R_P$ implies
$S_W \cdot S_P$ close to unity.
For $S_P$ chiral perturbation theory predicts a finite-size correction of
the form\cite{ONECPTH},
\begin{equation}
S_P = 1 + \frac{M_\pi^2}{24 F_\pi^2 ( M_\pi a L)^3 } \ .
\end{equation}
This formula indicates that the deviation of $S_P$ from unity would be less
than 1\% in our simulation.
Hence we conclude $S_W\approx 1$.
%
%
\subsection{ \label{OperatorMatching} Operator Matching }
For the quark field normalization we employ the KLM factor\cite{KLM}
\begin{equation}
  \psi^{continuum} =
  \sqrt{ 1 - \frac{ 3 \kappa }{ 4 \kappa_c } } \ \psi^{lattice} \ .
\end{equation}
Due to CPS symmetry the weak operator $Q_+$ defined in (\ref{QP}) does not
mix with other operators\cite{BSKpipi}.
With the tadpole improvement (with the factor $u_0=1/8\kappa_c$)
the multiplicative renormalization factor for $Q_+$, which relates
the lattice operator to the continuum one at a scale $\mu$,
is given by\cite{Z-M,Z-BS}
\begin{equation}
    Z(\mu) = 1 + { g^2_{\overline{\rm MS}} (\mu) \over 16\pi^2 }
       [ - 4 \log ( \mu a / \pi ) - 21.140 ]
\label{Z}
\end{equation}
where the naive dimensional regularization (NDR) is taken
with the modified minimum subtraction scheme ($\overline{\rm MS}$) in the continuum.

We employ the  $\overline{\rm MS}$ coupling constant estimated as follows.
First we obtain $g_V^2$ by\cite{TPI}
\begin{equation}
  - \log P = { 1 \over 3 } g_V^2(3.41/a)
       \{ 1 - ( 1.19 + 0.017 N_f ) \frac{ g_V^2 (3.41/a) }{ 4 \pi } + O(g_V^4) \}
\label{DEFOFAV}
\end{equation}
with $P$ the average plaquette.
Next we calculate $\Lambda_V$ from $g_V^2$ using
\begin{equation}
  \log ( \frac{ 3.41/a }{ \Lambda_V } )^2
         = \frac{ 1 }{ \beta_0 x }
         + \frac{ \beta_1 }{ \beta_0^2 }
           \log \frac{ \beta_1 x / \beta_0 }{ 1  +  \beta_1 x / \beta_0 }
\end{equation}
where $\beta_0 = 11 - 2 N_f / 3$, $\beta_1 = 102 - 38 N_f /3$,
and $x=g_V^2(3.41/a)/(4\pi)^2$.
A perturbative relation $\Lambda_{\overline{\rm MS}} = 0.6252 \cdot \Lambda_V$
then yields $\Lambda_{\overline{\rm MS}}$,
with which we can calculate $g_{\overline {\rm MS}}^2(\mu)$ at any scale $\mu$.
In the present calculation we find
$\Lambda_{\overline{\rm MS}} = 293(11){\rm MeV}$ with $P=0.605$
and $1/a=2.67(10){\rm GeV}$ at $\beta=6.1$.

Let $A_2$ be the physical amplitude for $\Delta I=3/2$ $K\to\pi\pi$ decay.
Experimentally,
\begin{equation}
\sqrt{ 3 \over 2 } \cdot {\rm Re} A_2
              \cdot [ \frac{ G_F }{\sqrt{2}} \cdot V_{us}^* V_{ud} ]^{-1}
 =  10.4 \times 10^{-3}\ \  {\rm GeV}^3 \ .
\end{equation}
The relation of the decay amplitude to the matrix element of $Q_+$ is
\begin{equation}
\sqrt{ 3 \over 2 } \cdot {\rm Re} A_2
              \cdot [ \frac{ G_F }{\sqrt{2}} \cdot V_{us}^* V_{ud} ]^{-1}
 = C_+^{(N_f)}(\mu) \cdot \langle \pi^+ \pi^0 | Q_+^{(N_f)}(\mu) | K^+ \rangle \ .
\end{equation}
On the right-hand side $C_+^{(N_f)}(\mu)$ and $Q_+^{(N_f)}(\mu)$ are
the Wilson coefficient function
and the renormalized weak operator at a scale $\mu$ with superscript
$N_f$ the number of quark flavors appropriate for the scale $\mu$.
We choose $\mu=2{\rm GeV}$ to estimate the physical amplitude,
and hence $N_f=4$.

In our calculation, matching of the lattice operator $Q_+^{lattice}$
to the continuum operator $Q_+^{(4)}(2{\rm GeV})$ is not straightforward
since the simulation is carried out in quenched QCD ($N_f=0$).
To treat this problem we proceed in the following way.
We first match the lattice operator to the continuum operator
$Q_+^{(0)}$ for $N_f=0$ at a scale $q^*$ using the renormalization
factor $Z(q^*)$ in (\ref{Z}): $Q_+^{(0)}(q^*)=Z(q^*)Q_+^{lattice}$.
The operator $Q_+^{(0)}(\mu)$ at any scale $\mu$ can then be obtained
by renormalization group evolution in the continuum:
\begin{eqnarray}
   Q^{(0)}_{+} ( \mu ) &=& U^{(0)}( \mu, q^* ) Q^{(0)}_{+} ( q^* )   \nonumber\\
                       &=& U^{(0)}( \mu, q^* ) Z(q^*) Q^{lattice}_{+}
\end{eqnarray}
where $U^{(N_f)}(\mu,\mu')$ is the two-loop
renormalization group running factor
from scale $\mu'$ to $\mu$ and it is given by
\begin{equation}
U^{(N_f)}(\mu,\mu')
 = \biggl( \frac{ g^2(\mu ) }{ g^2 (\mu' ) } \biggr)^{ \frac{ \gamma_0 }{ 2 \beta_0 } }
       \biggl[ 1 +  \frac{ g^2(\mu) - g^2(\mu') }{ 16 \pi^2 }
            \Bigl( \frac{ \gamma_1 \beta_0 - \gamma_0 \beta_1 }{ 2 \beta_0^2 }
            \Bigr)
       \biggr] \ .
\end{equation}
Here $\gamma_0 = 4$ and $\gamma_1=-7 + 4 N_f / 9$ are the one- and
two-loop anomalous dimensions for $Q_+$\cite{BJW}.

In the spirit of tadpole improvement, the matching point $q^*$
from the lattice to the continuum operator should be chosen to minimize
higher order contributions in the renormalization factor $Z(q^*)$.
Since an estimate of this value is not available, however,
we take $q^*=1/a$ or $\pi/a$ and investigate the $q^*$ dependence
of the decay amplitude.

We still need to relate the operator $Q^{(0)}_{+}$ of the $N_f=0$ theory 
to the operator $Q^{(4)}_{+}$ of the $N_f=4$ theory.  
Whether such a matching is possible is a problem generally encountered 
in quenched QCD calculations of weak matrix elements.  
As a working hypothesis, we assume that there is a scale $k^*$, 
typical of the $K^+\to\pi^+\pi^0$ process, 
at which the $N_f=0$ operator matches with the $N_f=4$ operator, 
\begin{equation}
   U^{(0)}( k^*, q^* ) Q^{(0)}_{+} (q^*) = Q^{(4)}_{+} ( k^* ) \ .
\label{KST}
\end{equation}
We then estimate the decay amplitude for the $N_f=4$ theory by
\begin{eqnarray}
& & C_+^{(4)}(\mu) \cdot \langle \pi^+ \pi^0 | Q_+^{(4)}(\mu) | K^+  \rangle  \cr
&=& C_+^{(4)}(\mu) \cdot U^{(4)}(\mu, k^*)
        \cdot \langle \pi^+ \pi^0 | Q_+^{(4)}(k^*) | K^+ \rangle \cr
&=& C_+^{(4)}(\mu) \cdot D(\mu, k^*, q^*)
         \cdot \langle \pi^+ \pi^0 | Q_+^{lattice} | K^+ \rangle
\end{eqnarray}
where
\begin{equation}
   D(\mu, k^*, q^*) = U^{(4)}( \mu, k^* ) U^{(0)}( k^*, q^*  ) Z(q^*) \ .
\label{eq:matching}
\end{equation}
For the renormalization group evolution in the continuum
we follow Buchalla {\it et al.}\cite{BURAS}.
In particular we use their $C_+^{(4)}(2{\rm GeV})=0.859$ with
$\Lambda^{(4)}_{\overline{\rm MS}}=215{\rm MeV}$.

The value of the matching scale $k^*$ is not known.  
The variation of $D(\mu, k^*, q^*)$ with respect to the scale 
$k^*$, however, arises from the difference of the $\Lambda$ parameter
and the anomalous dimension of $Q_+$ for $N_f=0$ and $4$, and so it is 
expected to be small.
The values of $D( \mu, k^*, q^*)$ for several $k^*$ 
are tabulated for $q^*=1/a$ and $\pi/a$ in Table~\ref{KS}.
We observe that the dependence on $k^*$ is indeed very small,  
and we set $k^*=1{\rm GeV}$ in the following analysis.

Let us note that the difference of $D( \mu, k^*, q^*)$
for $q^*=1/a$ and $\pi/a$ is about $10\%$.
This is the largest systematic error in our operator matching procedure
beside the assumption of the matching scale $k^*$,
and it is comparable to our statistical errors.
%
%
\section{\label{FOUR_S} Results for the $K^+\to\pi^+\pi^0$ amplitude }
%
%
\subsection{ Decay amplitude with tree-level CHPT }
As in the previous work\cite{BSKpipi,KPPE-GA} we take
degenerate strange and up-down quarks, and assume
all external mesons at rest.
The amplitude obtained with this kinematics is clearly unphysical,
having an energy injection at the weak operator.
In order to relate the lattice result to the physical amplitude
information is needed on the dependence of the amplitude on the $K$ and
$\pi$ masses away from the physical point.

Earlier calculations have used chiral perturbation theory (CHPT)
at tree level for this purpose.
The operator $Q_+$ is decomposed under chiral ${\rm SU(3)}_L$ into terms belonging to
$[ {\bf  8}, \Delta I=1/2 ]$,
$[ {\bf 27}, \Delta I=1/2 ]$ and
$[ {\bf 27}, \Delta I=3/2 ]$.
The $[ {\bf 27}, \Delta I=3/2 ]$ part of $Q_+$, which contributes to
$K^+\to\pi^+\pi^0$, is given by
\begin{equation}
\frac{1}{3} \cdot Q_4 = \frac{1}{3} \cdot
  [  2 \cdot Q_+
    -  \bar{s} \gamma_\mu ( 1- \gamma_5 ) d \ \bar{d} \gamma_\mu ( 1- \gamma_5 ) d ] \ .
\label{QFOUR}
\end{equation}
In general the ${\bf 27}$ operator in QCD can be described by operators
in CHPT as,
\begin{equation}
  {\cal O}_{27}^{{\rm QCD}} = \alpha_{27} \cdot R^{ij}_{kl} \cdot
     ( \Sigma \partial_\mu \Sigma^\dagger )_{ik}
     ( \Sigma \partial_\mu \Sigma^\dagger )_{jl}
\label{O27}
\end{equation}
where, for $Q_4$, the non-vanishing components of the tensor $R^{ij}_{kl}$
are
$R^{21}_{31} = R^{12}_{13} = R^{12}_{31} = R^{21}_{31} =  \frac{1}{2}$ and
$R^{22}_{32} = R^{22}_{23} = - \frac{1}{2}$ \ ,
and the pseudoscalar meson field is given by
\begin{equation}
   \Sigma = {\rm e}^{ i \pi / f }
\label{sigmaf}
\end{equation}
for the full theory, or
\begin{equation}
   \Sigma = {\rm e}^{ i \pi / f } {\rm e}^{ i \eta' / \sqrt{3} f }
\label{sigmaq}
\end{equation}
for the quenched theory.
At tree level of CHPT one obtains the formula connecting the
physical amplitude and that calculated on the lattice\cite{BSKpipi} :
\begin{equation}
  \langle \pi^+ \pi^0  | Q_+ | K^+  \rangle_{phys}
= {  { m_K^2 - m_\pi^2 } \over { 2 M_\pi^2 } }
     \cdot \biggl( \frac{ \alpha_{27} }{ \alpha_{27}^q } \biggr)
     \cdot \biggl( \frac{ f_q }{ f } \biggr)^3
     \cdot \langle \pi^+ \pi^0  | Q_+ | K^+  \rangle_{lattice}
\label{TREE}
\end{equation}
where $m_K=497{\rm MeV}$ and $m_\pi=136{\rm MeV}$ are physical masses,
and $M_\pi$ is the degenerate $K$ and $\pi$ masses on the lattice.
We emphasize that the constant $\alpha_{27}$ and the tree-level decay
constant $f$ may take different values in the full and quenched theories.
We denote the constants in quenched
theory with superscript $q$.

\begin{figure}
\centerline{\epsfxsize=14.0cm \epsfbox{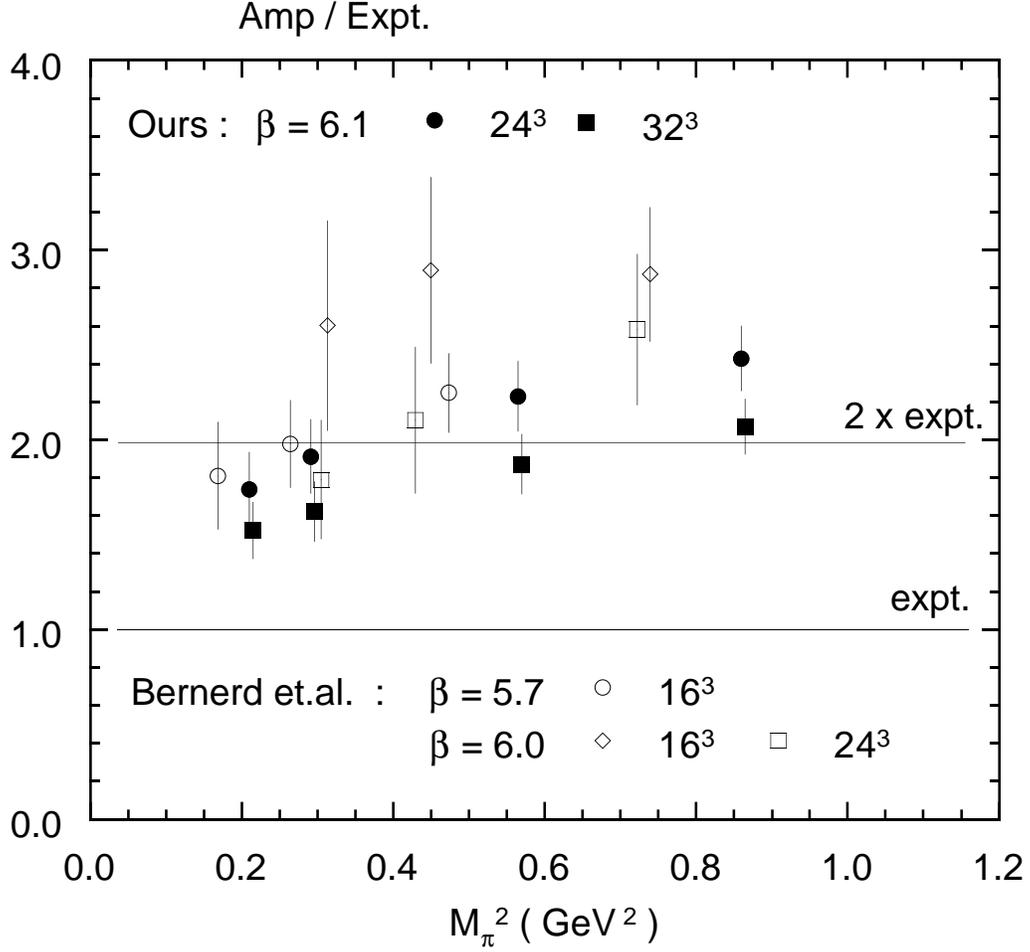}}
\caption{ \label{COMPARISON}
Comparison of our results for $C_+ \cdot \langle \pi^+ \pi^0 | Q_+ | K^+ \rangle$
normalized by the experimental value obtained with tree-level CHPT relation (\ref{TREE})
for $q^*=\pi/a$ at $\beta=6.1$ with those of previous work{\protect\cite{BSKpipi}}
at $\beta=6.0$ and $5.7$.
Results are plotted as a function of lattice meson mass $M_\pi^2$.
Traditional $\protect\sqrt{2\kappa}$ normalization is employed for quark fields
and tadpole-improvement is not applied in the renormalization factor.
}
\end{figure}

In Fig.~\ref{COMPARISON} we compare the decay amplitude
$C_+ \cdot \langle \pi^+ \pi^0 | Q_+ | K^+ \rangle$ of the previous work
at $\beta=5.7$ and $6.0$\cite{BSKpipi} with ours at $\beta=6.1$.
Here, as a working hypothesis, we set $\alpha_{27}$ and $f$
to be equal in full and quenched theories.
For the sake of comparison, our data are analyzed in a manner parallel
to that in Ref.~\cite{BSKpipi} as much as possible,
{\it i.e.,} employing the traditional $\sqrt{2\kappa}$ normalization for
quark fields, no tadpole improvement in the renormalization factor, and
applying the tree-level relation (\ref{TREE}).
The matching factor (\ref{eq:matching}) with $q^*=\pi/a$ is applied
in our results for consistency.
Since the normalization adopted in Ref.~\cite{BSKpipi} for comparison with
experiment differs from ours,
we plot the results divided by the experimental value.
In view of various differences in the simulation parameters and details of
analysis procedures, the values from the two studies are taken to be consistent,
both being larger than experiment roughly by a factor of two.

Let us note that our results, which attain errors of about 10\%, show a clear
dependence on the lattice meson mass $M_\pi$.
The presence of finite-size effects is also evident,
exhibiting a decrease between $24^3$ and $32^3$ spatial sizes.
We observe that these features were present in the previous simulations when
examined in the light of our results, but they were not evident at the time
because of large statistical errors of $20-30$\%.

\begin{figure}
\centerline{\epsfxsize=14.0cm \epsfbox{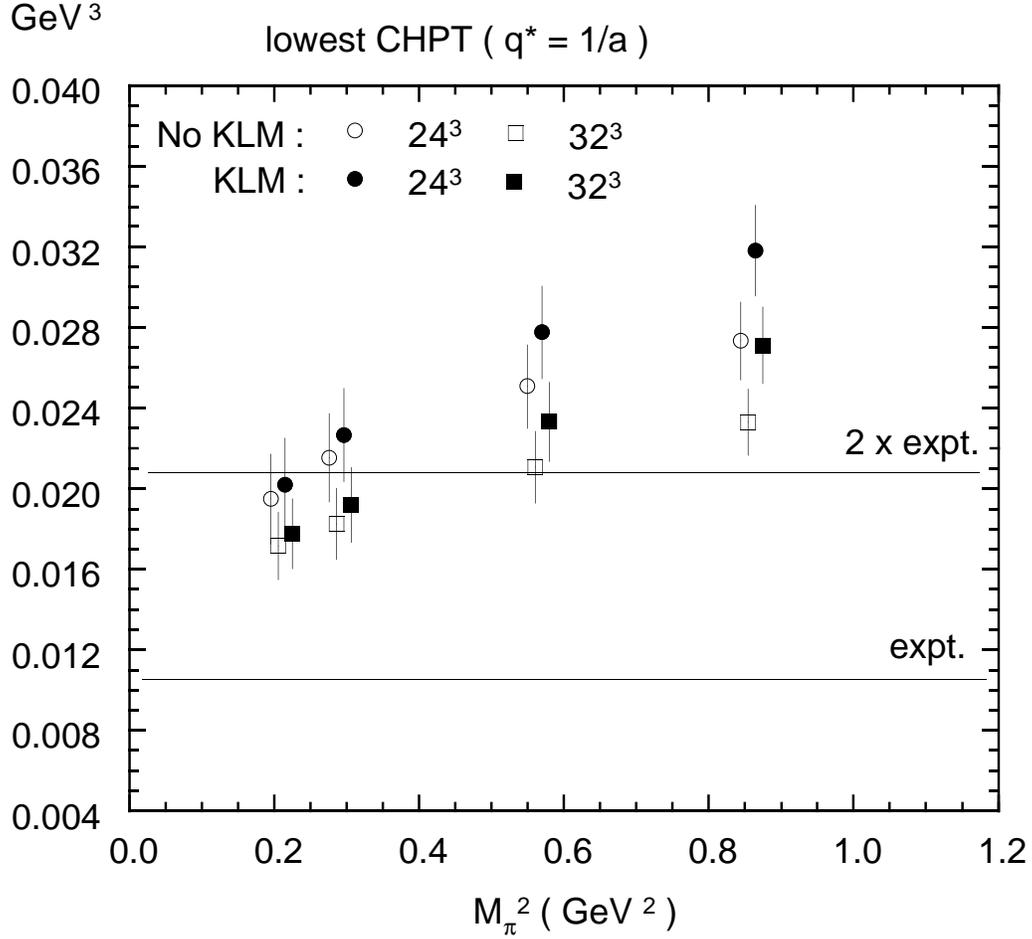}}
\caption{ \label{FTREE}
Decay amplitude $C_+ \cdot \langle \pi^+ \pi^0 | Q_+ | K^+ \rangle$ for $q^*=1/a$
as a function of lattice meson mass $M_\pi^2$ for tree-level CHPT.
Circles and squared refer to decay amplitude obtained on a $24^3$ and $32^3$ lattice.
Open symbols correspond to the traditional
$\protect\sqrt{2\kappa/2\kappa_c}/2$ normalization factor of Wilson quark fields,
while filled symbols are for the KLM normalization.
}
\end{figure}

In Fig.~\ref{FTREE} we show how the use of the KLM normalization affects
the meson mass dependence of the decay amplitude (plotted with filled symbols)
as compared to the conventional normalization
$\sqrt{2\kappa/2\kappa_c}/2$ (open symbols).
While the amplitudes for small $M_\pi$ change only slightly,
those for larger $M_\pi$
increase by about $20\%$, which is beyond the statistical error by a factor of two.
A significant meson mass dependence and finite-size effects observed in
our data show that tree-level CHPT is inadequate to extract the physical
amplitude from lattice calculations.
%
%
\subsection{ Decay amplitude with one-loop CHPT }
Recently Golterman and Leung have carried out a one-loop calculation of CHPT
for the decay amplitude in full and quenched QCD for degenerate
and non-degenerate $K$ and $\pi$ mesons\cite{ONECPTH}.
Their formula also includes finite-size correction terms.
Combining with the one-loop formula calculated for the physical point\cite{BK},
we analyze how our results are changed by one-loop effects of CHPT.

Let us denote by $\langle \pi^+ \pi^0 | Q_+ | K^+ \rangle_{phys}$ the physical amplitude
in the full theory with non-degenerate $K$ and $\pi$ mesons of mass $m_K$ and $m_\pi$,
and by $\langle \pi^+ \pi^0 | Q_+ | K^+ \rangle_{lattice}$
the amplitude in the quenched theory with degenerate $K$ and $\pi$ mesons of mass $M_\pi$.
According to Golterman and Leung,
\begin{equation}
   \langle \pi^+ \pi^0 | Q_+ | K^+ \rangle_{phys}
 = \frac{ m_K^2 - m_\pi^2 }{ 2 M_\pi^2}
     \cdot \biggl( \frac{ \alpha_{27} }{ \alpha_{27}^q } \biggr)
     \cdot \biggl( \frac{ f_q }{ f } \biggr)^3
     \cdot Y
     \cdot  \langle \pi^+ \pi^0 | Q_+ | K^+ \rangle_{lattice}
\label{ONELOOP}
\end{equation}
where $\alpha$ and $f$ are defined in (\ref{O27}--\ref{sigmaq}),
and the factor $Y$ is given by
\begin{equation}
  Y = \frac{  1 + \frac{ m_\pi^2 }{ (4 \pi f_\pi )^2 } \Bigl[ U + d \Bigr] }
           {  1 + \frac{ M_\pi^2 }{ (4 \pi F_\pi )^2 }
    \Bigl[
       - 3 \log \Bigl( \frac{ M_\pi }{\Lambda^q} \Bigr)^2 + F( M_\pi a L ) + d_q
    \Bigr]
              } \ .
\label{Y}
\end{equation}
The numerator of $Y$ represents the one-loop effect in the full theory,
and the denominator is the corresponding effect in the quenched theory.
The dimensionless constants $d$ and $d_q$ are the contact term coefficients
arising from the $O(p^4)$ terms of the chiral Lagrangian.
$f_\pi$ and $F_\pi$ are the one-loop corrected
decay constants in the full and quenched theories,
which differ from the tree-level values $f$ and $f_q$.
In the numerator of $Y$, $U$ is a complicated function of physical $K$ and $\pi$ masses,
the decay constant $f_\pi$ and $f_K$, and the cutoff of CHPT for the full
theory $\Lambda^{cont}$, and a numerical approximation is
\begin{equation}
    U = A  + B \cdot \log \Bigl( \frac{ m_\pi }{ \Lambda^{cont} } \Bigr)^2
\label{U}
\end{equation}
where $A=-104.73$ and $B=-29.57$ for $m_\pi=136{\rm MeV}$, $m_K=497{\rm MeV}$,
$f_\pi=132{\rm MeV}$, and $f_K=160{\rm MeV}$.
In the denominator of (\ref{Y}), $\Lambda^q$ is the cutoff of CHPT for quenched QCD,
and $F( M_\pi a L)$ represents finite-size corrections for a spatial size $L$
which takes the form
\begin{equation}
  F( M_\pi a L ) = \frac{ 17.827 }{ M_\pi a L }
                 + \frac{ 12 \pi^2 }{ ( M_\pi a L )^3 } \ .
\label{finitL}
\end{equation}
We set $\alpha_{27}$ and $f$ to be equal in the quenched and full theories
as in the analysis with the tree-level CHPT.
We initially ignore the effects of $O(p^4)$ terms of the chiral Lagrangian 
$d$ and $d_q$. 
We leave $\Lambda^{cont}$ and $\Lambda^q$ to be different, however, 
and examine the dependence of the results on these cutoffs.

\begin{figure}
\centerline{\epsfxsize=14.0cm \epsfbox{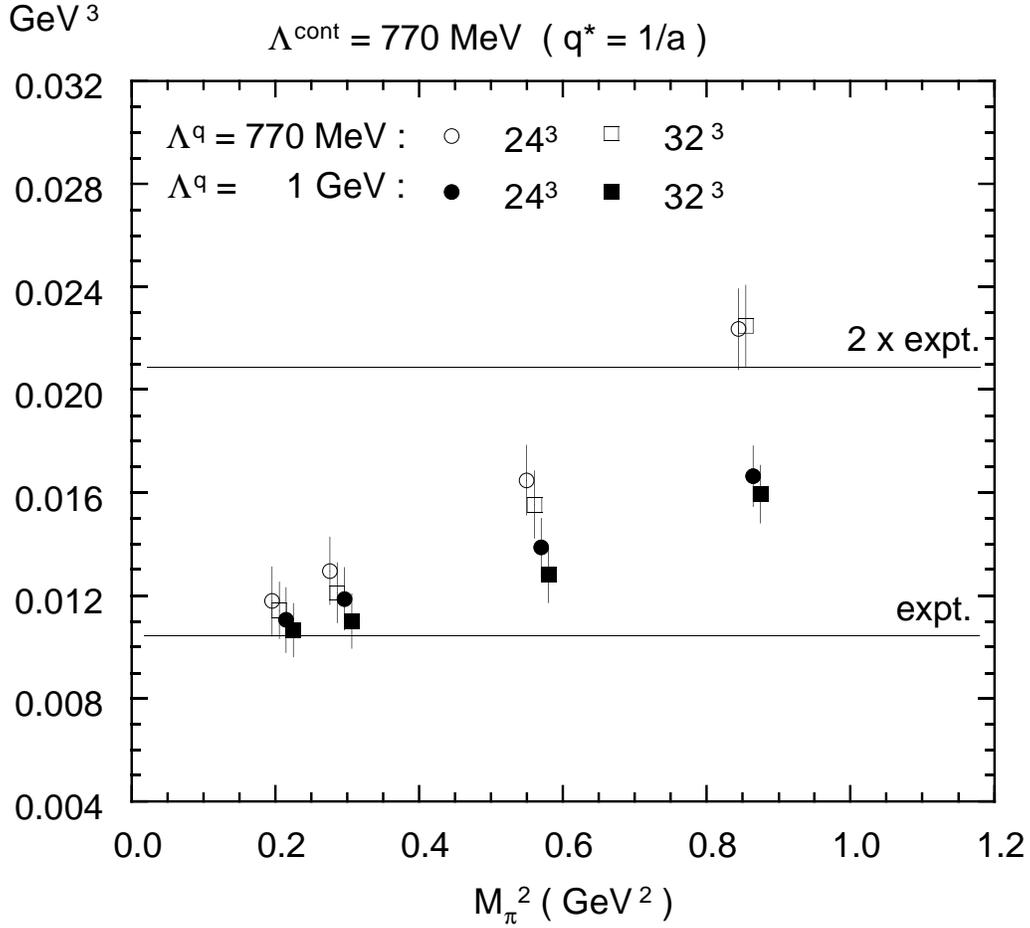}}
\caption{ \label{FRHO}
Decay amplitude $C_+ \cdot \langle \pi^+ \pi^0 | Q_+ | K^+ \rangle$ for $q^*=1/a$ obtained with
one-loop CHPT for $\Lambda^{cont}=770{\rm MeV}$ plotted as a function of $M_\pi^2$.
Circles and squares refer to data for $24^3$ and $32^3$ spatial sizes.
Open symbols are for $\Lambda^{q}=770{\rm MeV}$ and filled symbols for
$\Lambda^{q}=1{\rm GeV}$.
}
\end{figure}

\begin{figure}
\centerline{\epsfxsize=14.0cm \epsfbox{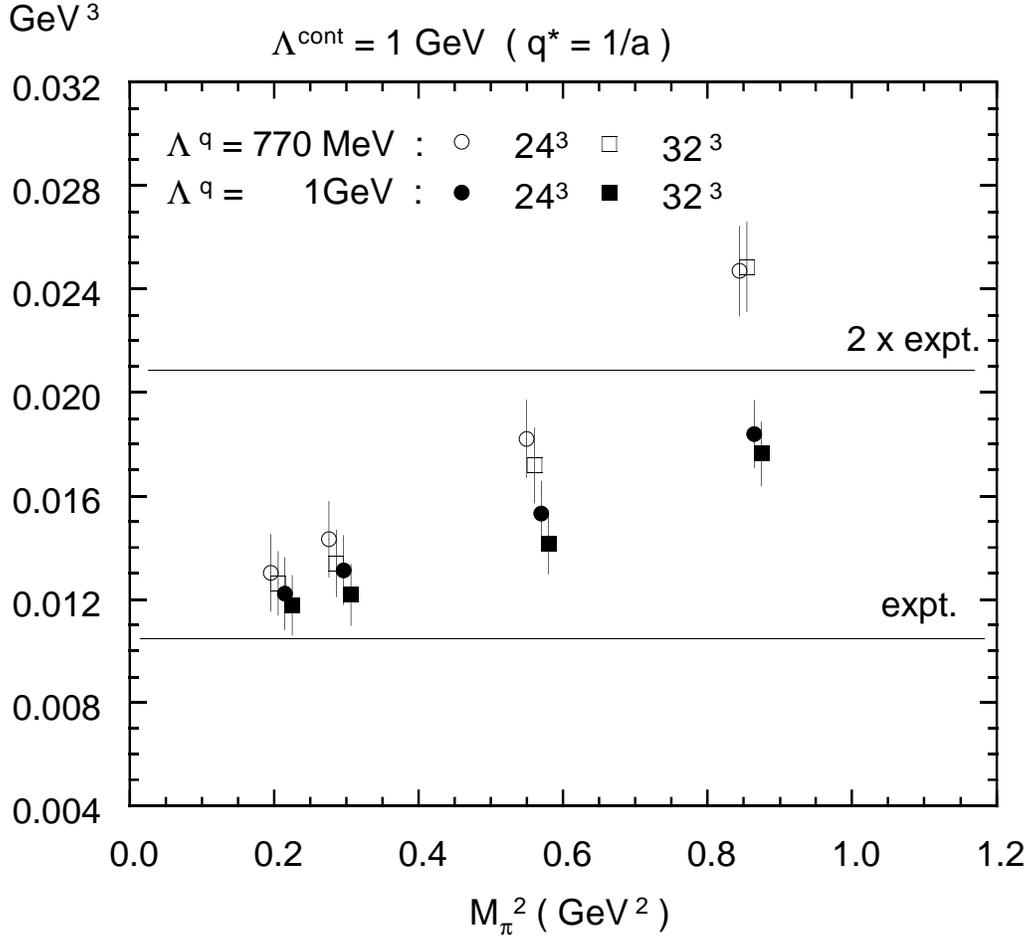}}
\caption{ \label{FO}
Decay amplitude $C_+ \cdot \langle \pi^+ \pi^0 | Q_+ | K^+ \rangle$ for $q^*=1/a$
as a function of $M_\pi^2$ obtained with one-loop CHPT for $\Lambda^{cont}=1{\rm GeV}$.
Meaning of symbols are the same as in Fig.~\ref{FRHO}.
}
\end{figure}

In Fig.~\ref{FRHO} we plot the one-loop corrected decay
amplitude for $\Lambda^{q}=770{\rm MeV}$ and $1{\rm GeV}$
for the choice $\Lambda^{cont}=770{\rm MeV}$.
We set $f_\pi=F_\pi = 132{\rm Mev}$ in (\ref{Y}), and the finite-size
corrections  $F( M_\pi a L)$ are taken into account.
The results of a similar analysis for the choice $\Lambda^{cont}=1{\rm GeV}$
are plotted in Fig.~\ref{FO}.

An important feature observed in Figs.~\ref{FRHO} and \ref{FO} 
is that the size dependence seen with the tree-level analysis in Fig.~\ref{FTREE}
is removed after finite-size corrections at the one-loop level.
At the same time, the amplitude decreases by $30-40$\% over 
the range of meson mass covered in our simulation.

Another noteworthy feature in Figs.~\ref{FRHO} and \ref{FO}
is that a sizable lattice meson mass dependence still remains in 
the amplitude, and that the magnitude of the slope depends
sensitively on the choice of $\Lambda^{q}$.
This feature can be understood as arising from the $O(p^4)$ 
coupling constants in the quenched theory, {\it i.e.,} $d_q$ in (\ref{Y}), 
which was ignored above.
If we denote our present results by
$\langle \pi^+ \pi^0 | Q_+ | K^+ \rangle_{ours}$
we find from (\ref{Y}) that
\begin{equation}
   \langle \pi^+ \pi^0 | Q_+ | K^+ \rangle_{ours}
 = \langle \pi^+ \pi^0 | Q_+ | K^+ \rangle_{phys}
 \cdot
   \frac{1 + \frac{ M_\pi^2 }{ (4 \pi F_\pi )^2 } d_q }
        {1 + \frac{ m_\pi^2 }{ (4 \pi f_\pi )^2 } d   }
\label{CONTACT}   
\end{equation}
showing the presence of a term linear in $M_\pi^2$. 
We note furthermore that $d_q$ actually depends on $\Lambda^q$:  
$d_q=d_q(\Lambda^q)$. 
Since the total $O(p^4)$ correction in the denominator of (\ref{Y})
should be independent of the cutoff $\Lambda^q$, 
$d_q$ for different values of $\Lambda^q$ varies according to 
\begin{equation}
 d_q(\Lambda^q) = d_q(\Lambda'^q) 
         - 3 \log \left( \frac{\Lambda^q}{\Lambda'^q} \right)^2 \ .
\label{CONV}
\end{equation}

To compare these relations with our results,
we fit $\langle \pi^+ \pi^0 | Q_+ | K^+ \rangle_{ours}$ 
as a function of $M_\pi^2$ to the form (\ref{CONTACT}). 
Employing data for which the value of $M_\pi$ does not exceed the cutoff, 
we find $d_q(1{\rm GeV})/(4\pi)^2 \approx 0.015$ 
and $d_q(770{\rm MeV})/(4\pi)^2\approx 0.025$.
The difference 
$d_q(1{\rm GeV})/(4\pi)^2-d_q(770{\rm MeV})/(4\pi)^2\approx -0.01$ 
is in good agreement with the value 
$-3\log\left(1{\rm GeV}/770{\rm MeV}\right)^2/(4\pi)^2=-0.0099$ 
expected from (\ref{CONV}). 
These results show that the uncertainties associated with $d_q$
can be removed by a chiral extrapolation of our amplitude
to the chiral limit $M_\pi=0$.

A further consequence of (\ref{CONTACT}) is that a correction
due to the $O(p^4)$ term $d$ in the full theory remains even after 
taking the limit $M_\pi\to 0$ in our results. 
In order to examine the magnitude of this uncertainty,
we use an estimate 
$d(\Lambda^{cont})/(4\pi)^2 = 0.003(14)$ at $\Lambda^{cont}=m_\eta$ 
from a phenomenological analysis\cite{BK}.
In view of the formula
\begin{equation}
     d(\Lambda^{cont}) 
  =  d(\Lambda'^{cont}) 
     - 29.57 \cdot \log \left( \frac{\Lambda^{cont}}{\Lambda'^{cont}} \right)^2
\label{CONVFULL}
\end{equation}
obtained from (\ref{Y}) and (\ref{U}), this leads to a value 
$d(770{\rm MeV})/(4\pi)^2 \approx -0.12$ and
$d(  1{\rm GeV})/(4\pi)^2 \approx -0.22$.
These values imply that the physical decay amplitude is $10$\% lower 
than our results for $\Lambda^{cont}=770{\rm MeV}$, and $20$\% 
for $\Lambda^{cont}=1{\rm GeV}$.
This provides an explanation of a discrepancy of about 10\% observed
in Figs.~\ref{FRHO} and \ref{FO} 
between the values of $\langle \pi^+ \pi^0 | Q_+ | K^+ \rangle_{ours}$ 
calculated with $\Lambda^{cont}=770{\rm MeV}$ and $1{\rm GeV}$. 
Let us add a remark that the values of $d$ estimated above 
for $\Lambda^{cont}=770{\rm MeV}$ and $1{\rm GeV}$ is an order 
of magnitude larger compared to those of $d_q$ for the quenched theory. 

We find from this analysis that including the correction due to 
the $O(p^4)$ coupling constants is possible
if an accurate value of $d$ is known from phenomenological studies. 
Since this is not yet the case\cite{BK}, 
we shall not pursue this point further here, leaving the correction 
as a source of uncertainty in our final results. 

The amplitude obtained from 
$\langle \pi^+ \pi^0 | Q_+ | K^+ \rangle_{ours}$ by a chiral extrapolation 
to the limit $M_\pi=0$ is listed in Table~\ref{FINALFIT}
for several choices of the cutoff and the operator matching point $q^*$.
In the results in Table~\ref{FINALFIT},
the systematic error due to the matching scale $q^*$ is about 10\%. 
Statistical errors are larger (about 20\%),
mainly due to a linear extrapolation to the chiral limit.
Within these uncertainties and that of 10--20\% due to the $d$ term discussed above, 
the values in Table~\ref{FINALFIT}
are consistent with the experiment $10.4\times 10^{-3}{\rm GeV}^3$.
%
%
\section{\label{FIVE_S} $B_K$ from the $K^+ \to \pi^+ \pi^0$ amplitude }
The $\Delta S=2$ four-quark operator defined by
\begin{equation}
O_{\Delta S=2} =
  \bar{s} \gamma_\mu ( 1 - \gamma_5 ) d
\ \bar{s} \gamma_\mu ( 1 - \gamma_5 ) d
\end{equation}
belongs to the same ${\bf 27}$ representation as the operator $Q_4$
which is the $[ {\bf 27}, \Delta I=3/2 ]$ part of $Q_+$.
As a consequence one can obtain the $K$ meson $B$ parameter $B_K$
from the $K^+ \to \pi^+ \pi^0$ amplitude using CHPT.
The non-vanishing component of the tensor $R^{ij}_{kl}$ in (\ref{O27})
for this operator is given by $R^{22}_{33} = 1$.

The one-loop relation in CHPT for quenched QCD for the unphysical degenerate case
has been obtained by Golterman and Leung\cite{ONECPTH},
\begin{equation}
  B_K = { 1 \over { { 8 \over 3 } F_K } } \cdot
        {   {   3 \cdot \langle \pi^+ \pi^0 | Q_+ | K^+ \rangle } \over
            { { 3 \over \sqrt{2} } \cdot M_\pi^2 ( 1 + R + d_q ) }
        }
\label{BKCHPT}
\end{equation}
with the one-loop correction $R$ given by
\begin{equation}
  R = \frac{ M_\pi^2 }{ ( 4\pi F_\pi )^2 }
        \Bigl[ 3 \log \big( \frac { M_\pi}{ \Lambda^q } \big)^2
               + F( M_\pi a L )
        \Bigr] \ .
\label{BKCHPTR}
\end{equation}
Here $F_K$ and $F_\pi$ denote the $K$ and $\pi$ meson decay constants in
the quenched theory, and the other notations are the same as those
in (\ref{Y}--\ref{finitL}).

\begin{figure}
\centerline{\epsfxsize=14.0cm \epsfbox{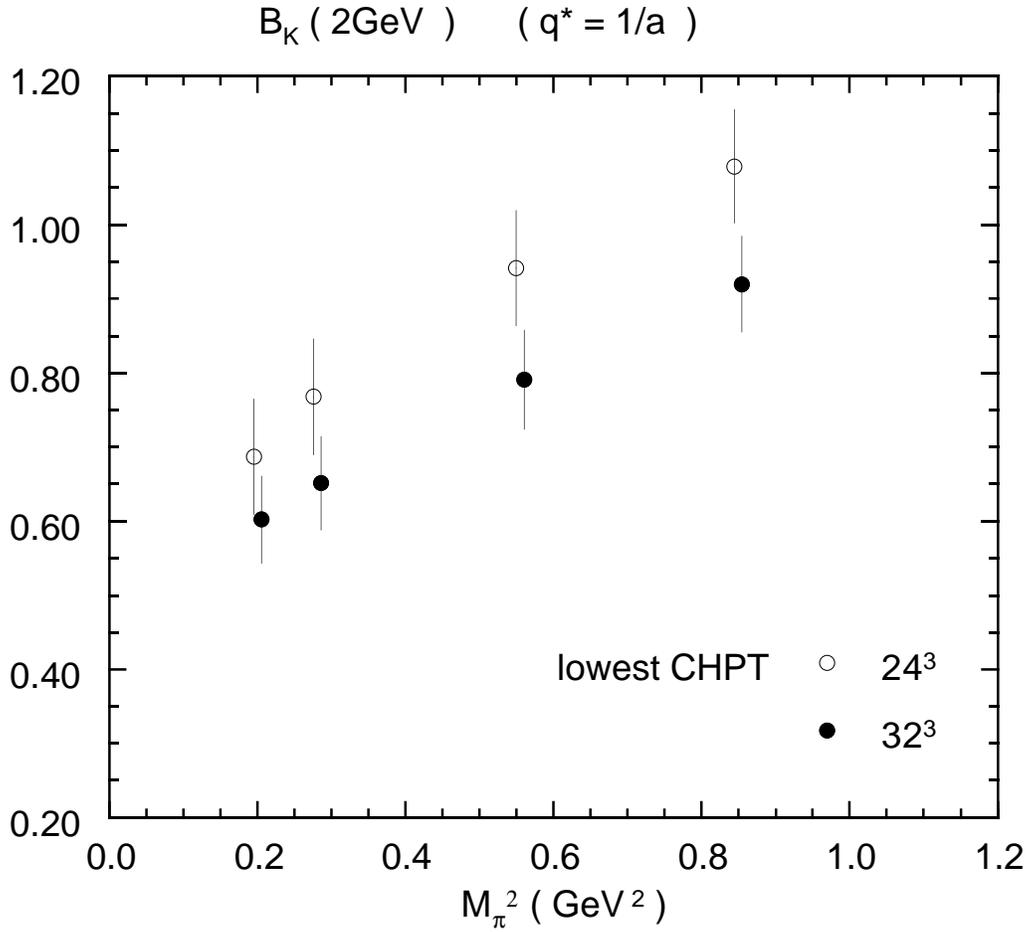}}
\caption{ \label{BKTREE}
$B_K(2{\rm GeV})$ for $q^*=1/a$ obtained from $K^+ \to \pi^+ \pi^0$ decay amplitude
as a function of $M_\pi^2$ obtained with tree level CHPT.
Circles and squares refer to data for $24^3$ and $32^3$ spatial sizes.
}
\end{figure}

\begin{figure}
\centerline{\epsfxsize=14.0cm \epsfbox{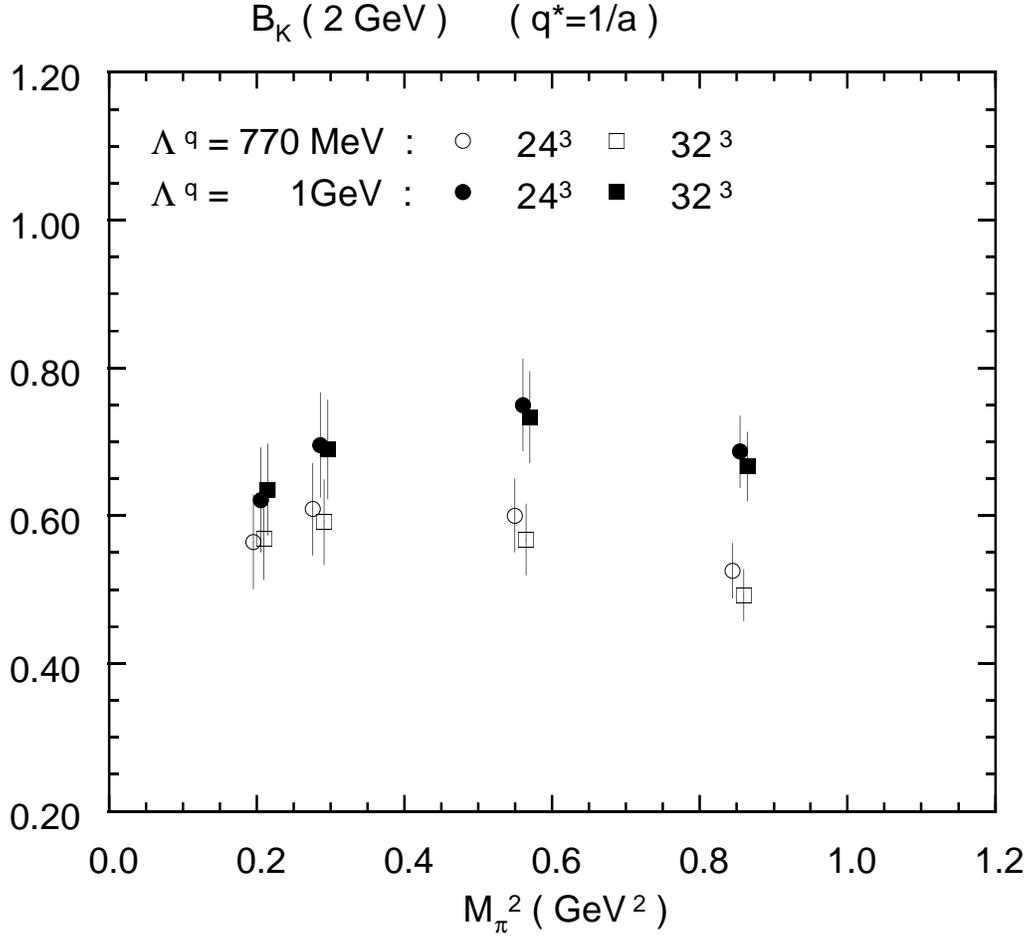}}
\caption{ \label{BKONE}
$B_K(2{\rm GeV})$ for $q^*=1/a$ obtained from $K^+ \to \pi^+ \pi^0$ decay amplitude
as a function of $M_\pi^2$ obtained with one-loop CHPT for $\Lambda^{q}=770{\rm MeV}$
and $1{\rm GeV}$.
Circles and squares refer to data for $24^3$ and $32^3$ spatial sizes.
Open symbols are for $\Lambda^{q}=770{\rm MeV}$ and filled symbols for
$\Lambda^{q}=1{\rm GeV}$.
}
\end{figure}

Our procedure for calculating $B_K$ from (\ref{BKCHPT}) is essentially
the same as for the $K^+ \to \pi^+ \pi^0$ amplitude
including the operator matching procedure,
although the coefficient function $C_+$ is absent in the present case.
In Figs.~\ref{BKTREE} and ~\ref{BKONE} we plot $B_K(2{\rm GeV})$
obtained from the $K^+ \to \pi^+ \pi^0$ decay amplitude with tree and one-loop CHPT.
We set $F_K=160{\rm MeV}$ and $F_\pi=132{\rm MeV}$ in (\ref{BKCHPT}) and (\ref{BKCHPTR}).
The one-loop CHPT effect and the cutoff dependence for a small $M_\pi^2$ region
are small compared with those for the decay amplitude.
At the physical $K$ meson mass $M_\pi^2=0.246{\rm GeV}^2$
$B_K$ takes almost the same value for different choices of
$\Lambda^{\rm q}$ and the lattice size.
In Table~\ref{BKtable} the average of the two data points with the
smallest $M_\pi$ is tabulated.
Our results, $B_K = 0.581(56)-0.663(67)$ are consistent with the
JLQCD value $B_K({\rm 2GeV}) = 0.68(11)$\cite{KURA} obtained at the same
coupling constant $\beta=6.1$ through a calculation of the
$K^0-\overline{K}^0$ matrix element of the $\Delta S=2$ operator
employing chiral Ward identities for determining the mixing coefficients.

A direct calculation of $B_K$ with the Wilson quark action has the complication
that the operator mixing problem of the $\Delta S = 2$ operator
has to be solved non-perturbatively, which causes large statistical errors.
In contrast, the $Q_+$ operator does not mix with other operators
as mentioned in section~\ref{OperatorMatching}.
Therefore, statistical errors of $B_K$ obtained from the $K^+ \to \pi^+ \pi^0$
amplitude is smaller.
Theoretical uncertainties associated with the use of CHPT, however,
are large in this approach that offsets the advantage of the present method.
In any case, our calculation, albeit with a significant error,
provides an independent check for $B_K$ for the Wilson quark action obtained
with the chiral Ward identity procedure, and also supports the validity of CHPT.
%
%
\section{\label{SIX_S} Conclusions }
In this article we have reported results of a study 
of the $K^+\to\pi^+ \pi^0 $ decay amplitude in quenched lattice QCD.
With a set of high statistics simulations we have found that the results show 
sizable finite-size effects, which, however, are consistent with those 
predicted by a recent one-loop calculation of CHPT. 
We have furthermore seen that a meson mass dependence which remains after 
inclusion of the one-loop corrections of CHPT in the prediction for the decay 
amplitude is due to effects of the $O(p^4)$ contact terms in the quenched theory.
Making an extrapolation to the chiral limit to remove these effects, 
we have found $8.9(1.7)\times 10^{-3} - 11.4(1.5)\times 10^{-3}{\rm GeV}^3$
for the physical value of the decay amplitude, depending on the choice of 
the cutoff parameter of CHPT.  
These values are consistent with experiment ($10.4\times 10^{-3}{\rm GeV}^3$).

The present result may be compared to those of the previous 
studies\cite{BSKpipi,KPPE-GA}
which gave decay amplitudes roughly twice larger than experiment.
Our smaller value originates from the two effects, one-loop corrections
as also noted by Golterman and Leung in their reanalysis of the old results,
and a decrease of the amplitude toward smaller values of $M_\pi$.

As a further application of the one-loop formula, we have calculated the
$B_K$ parameter, and found that it is consistent with a recent direct
calculation for $K^0-\overline{K}^0$ mixing.

The encouraging results we have obtained, however, should be taken
with several reservations.
The value of the $K^+\to\pi^+ \pi^0 $ decay amplitude 
estimated in the chiral limit suffers from uncertainties of 10--20\% 
due to the $O(p^4)$ contact terms of the full theory, 
because the phenomenological estimate available is not very accurate. 
A sizable finite-size correction of $30-40$\%,
while consistent with the one-loop prediction of CHPT,
raises the question whether ignoring higher order corrections can be justified.
Furthermore, various constants of CHPT, in particular the 
coefficient $\alpha_{27}$,
may differ between the quenched and full theories,
and we have no way of estimating or correcting the difference.
Reliability of CHPT for calculating unphysical amplitudes could also be an issue. 
Reducing these sources of uncertainties, especially those related to 
quenching and better controlling finite-size effects
would require a difficult task of carrying out simulations 
in full QCD on a physically large lattice. 
%
%
\section*{Acknowledgment}
We thank Maarten Golterman for informative correspondence on his results 
for the decay amplitude from chiral perturbation theory. 
This work is supported by the Supercomputer Project (No.~97-15) of High Energy
Accelerator Research Organization (KEK), and also in part by the Grants-in-Aid
of the Ministry of Education (Nos. 08640349, 08640350, 08640404,
09246206, 09304029, 09740226).
%
%

%
%
%
%
%
%
%
%
%
%
%
%
%
%
%
%
%
%
\begin{table}[t]
\begin{center}
\begin{tabular}{lllll}
$\kappa$  &  $aM_\pi$       &   $a \Delta $   & \multicolumn{2}{c}{$\langle \pi^+ \pi^0 | Q_+ | K^+ \rangle$}  \\
          &                 &   $(10^{-3} )$  &  from $R_W$    &  from $R_P$   \\
\hline
$L=24$ \\
$0.1520$  &  $0.3440(14)$  &   $7.9(1.7)$   &  $0.261(20) $  &  $0.271(19) $ \\
$0.1530$  &  $0.2776(17)$  &   $8.6(2.0)$   &  $0.151(14) $  &  $0.160(13) $ \\
$0.1540$  &  $0.1967(19)$  &   $9.3(2.9)$   &  $0.0617(81)$  &  $0.0680(69)$ \\
$0.1543$  &  $0.1653(21)$  &   $8.6(3.6)$   &  $0.0382(58)$  &  $0.0434(49)$ \\
\hline
$L=32$ \\
$0.1520$  &  $0.3459(10)$  &   $3.6(1.4)$   &  $0.229(17) $  &  $0.234(16) $ \\
$0.1530$  &  $0.2784(11)$  &   $4.2(1.5)$   &  $0.132(13) $  &  $0.135(11) $ \\
$0.1540$  &  $0.1914(13)$  &   $5.7(2.1)$   &  $0.0565(71)$  &  $0.0573(55)$ \\
$0.1543$  &  $0.1651(15)$  &   $7.1(2.8)$   &  $0.0383(55)$  &  $0.0380(37)$ \\
\end{tabular}
\end{center}
\caption{\label{RWRPT}
The mass shift $\Delta=M_{\pi\pi}-2M_\pi$ and
$\langle \pi^+ \pi^0 | Q_+ | K^+ \rangle$ from $R_W$ and $R_P$.
Here we assume $S_W = S_P = 1$.
These value are obtained by a single exponential fit over $t=18-46$ for $R_W$
and by a constant fit over $t=22-42$ for $R_P$.
}
\end{table}
%
%
\begin{table}[t]
\begin{center}
\begin{tabular}{lll}
 $k^* ( {\rm GeV} )$  & $q^* = 1/a$ & $q^* = \pi/a$ \\
\hline
  0.700 &     0.759038  & 0.830913    \\
  1.000 &     0.761556  & 0.833670    \\
  1.500 &     0.765198  & 0.837657    \\
  2.000 &     0.768126  & 0.840863    \\
\end{tabular}
\end{center}
\caption{\label{KS}
Values of $D( 2{\rm GeV}, k^*, q^*)$ for $\Lambda^{(4)}_{\overline{\rm MS}}=215{\rm MeV}$ and
$\Lambda^{(0)}_{\overline{\rm MS}}=293{\rm MeV}$.}
\end{table}
%
%
\begin{table}[t]
\begin{center}
\begin{tabular}{llllll}
&
& \multicolumn{4}{c}{$C_+ \cdot \langle \pi^+ \pi^0 | Q_+ | K^+ \rangle (\times 10^{-3} {\rm GeV}^3)$}\\
 $\Lambda^{cont}$   &  $\Lambda^{q}$   &  \multicolumn{2}{c}{$24^3$} & \multicolumn{2}{c}{$32^3 $} \\
 (GeV)              &  (GeV)           &  $q^*=1/a$  & $q^*=\pi/a$   & $q^*=1/a$    & $q^*=\pi/a$  \\
\hline
0.77  &  0.77  &  $ 9.3(1.9)$ & $10.2(2.1)$  & $8.9(1.7)$ & $ 9.7(1.9)$ \\
0.77  &  1.0   &  $ 9.4(1.3)$ & $10.3(1.4)$  & $8.8(1.1)$ & $ 9.6(1.2)$ \\
1.0   &  0.77  &  $10.3(2.1)$ & $11.3(2.3)$  & $9.8(1.9)$ & $10.7(2.1)$ \\
1.0   &  1.0   &  $10.4(1.4)$ & $11.4(1.5)$  & $9.7(1.2)$ & $10.6(1.3)$ \\
%
\end{tabular}
\end{center}
\caption{\label{FINALFIT}
Results of linear extrapolation of $ C_+ \cdot \langle \pi^+ \pi^0 | Q_+ | K^+ \rangle$
to $M_\pi^2=0$.
For $\Lambda^{q}=770{\rm MeV}$ fits are made with three points with
smaller $M_\pi$ as $M_\pi$ of the forth point exceeds the cutoff.
Statistical and extrapolation errors are combined.
The experimental value is $10.4\times 10^{-3}{\rm GeV}^3$.
}
\end{table}
%
%
\begin{table}[t]
\begin{center}
\begin{tabular}{llll}
       & tree        &  $\Lambda^q=0.77{\rm GeV}$  & $\Lambda^q=1{\rm GeV}$  \\
\hline
$24^3$ & $0.728(78)$ &  $0.587(64)$                 & $0.659(71)$             \\
$32^3$ & $0.627(63)$ &  $0.581(58)$                 & $0.663(67)$             \\
\end{tabular}
\end{center}
\caption{ \label{BKtable}
$B_K(2{\rm GeV} )$ at physical $K$ meson mass $M_\pi = 496{\rm MeV}$
obtained from the $K^+ \to \pi^+ \pi^0$ amplitude.
The row ``tree'' refers to result with lowest CHPT and others are obtained by one-loop CHPT for
$\Lambda^q=0.77{\rm GeV}$ and $1{\rm GeV}$.
}
\end{table}
%
%
\end{document}